\documentclass[aps,pra,twocolumn,amsmath,amssymb,footinbib,showpacs]{revtex4-1}

\newcommand{\Jnature}{Nature (London)}
\newcommand{\Jnatphys}{Nat. Phys.}

\newcommand{\Jprl}{Phys. Rev. Lett.}
\newcommand{\Jpr}{Phys. Rev.}
\newcommand{\Jpra}{Phys. Rev. A}
\newcommand{\Jprb}{Phys. Rev. B}

\newcommand{\Jrmp}{Rev. Mod. Phys.}

\newcommand{\Jepl}{Europhys. Lett.}
\newcommand{\Jnjp}{New J. Phys.}

\newcommand{\Jepjd}{Eur. Phys. J. D}

\newcommand{\Jjetp}{Sov. Phys. JETP}

\newcommand{\Jstatmech}{J. Stat. Mech.}

\newcommand{\Jphysrep}{Phys. Rep.}
\newcommand{\JRepProgPhys}{Rep. Prog. Phys.}

\newcommand{\JjphysA}{J. Phys. A: Math. Theor.}

\newcommand{\Jannphys}{Ann. Phys.}

\usepackage[english]{babel}
\usepackage{latexsym}
\usepackage{graphics}
\usepackage{subfigure}
\usepackage{epsfig}
\usepackage{color}
\usepackage{hyperref}
\hypersetup{
colorlinks=true,
citecolor=blue,
linkcolor=red,
urlcolor=black
}

\newcommand{\epsfboxmod}[1]{\epsfbox{#1}}

\newcommand{\infigbis}[2]{          \mbox{ \epsfxsize #1 \epsfboxmod{#2}}
                                      \vspace{-0.8cm}}

\newcommand{\ie}{{i.e.}}
\newcommand{\eg}{{e.g.}}

\newcommand{\e}{\textrm{e}}

\newcommand{\nc}{n_{\textrm{c}}}

\newcommand{\pt}{\mathbf{r}}

\newcommand{\Vr}{V_\textrm{\tiny R}}
\newcommand{\sigmar}{\sigma_\textrm{\tiny R}}
\newcommand{\Vrand}{V}

\newcommand{\Vscreen}{\mathcal{V}_{\varepsilon}}
\newcommand{\kmax}{k_{\varepsilon}^\textrm{max}}
\newcommand{\kmin}{k_{\varepsilon}^\textrm{min}}

\newcommand{\diffB}{D_\textrm{\tiny B}}
\newcommand{\lB}{l_\textrm{\tiny B}}
\newcommand{\Lloc}{L_\textrm{\tiny loc}}

\begin{document}

\title{Propagation of collective pair excitations in disordered Bose superfluids}

\author{Samuel Lellouch}
\author{Lih-King Lim}
\author{Laurent Sanchez-Palencia}
\affiliation{
  Laboratoire Charles Fabry,
  Institut d'Optique, CNRS, Univ Paris Sud 11,
  2 avenue Augustin Fresnel,
  F-91127 Palaiseau cedex, France
}

\date{\today}

\begin{abstract}
We study the effect of disorder on the propagation of collective excitations in a disordered Bose superfluid.
We incorporate local density depletion induced by strong disorder at the meanfield level,
and formulate the transport of the excitations in terms of a screened scattering problem.
We show that the competition of disorder, screening, and density depletion
induces a strongly non-monotonic energy dependence of the disorder parameter.
In three dimensions, it results in a rich localization diagram with four
different classes of mobility spectra,
characterized by either no or up to three mobility edges.
Implications on experiments with disordered ultracold atoms are discussed.
\end{abstract}


\maketitle

\section{Introduction}
The dynamics of correlated quantum systems attracts a growing attention sparked by
the recent development of quantum devices with long coherence times and dynamical control of parameters,
\eg~superconducting circuits and ultracold atoms~\cite{NaturePhysicsInsight2012}.
An additional asset of the latter is that disorder may be introduced in a controlled way~\cite{lsp2010,*modugno2010,*shapiro2012}.
Disorder may strongly affect dynamical processes,
mainly due to Anderson localization~\cite{anderson1958}.
Understanding the interplay of disorder and interactions in dynamical quantum systems is thus of fundamental
importance and localization in quantum systems is still the subject of active research~\cite{basko2006,bilas2006,oganesyan2007,lugan2007b,aleiner2010,pal2010,lugan2011,scarola2015}.
This topic has also been addressed in the context of quasi-periodic systems~\cite{michal2014,lellouch2014}.

In correlated quantum systems, most basic dynamical processes
are determined by the transport properties of their collective excitations~\cite{polkovnikov2011}.
An important starting point in the understanding of localization in correlated systems
thus relies on classification according to the symmetries of their excitations~\cite{evers2008}.
For Fermi systems, it is mostly based on the three classes of random matrices~\cite{beenakker1997}
as well as chiral or particle-hole symmetries~\cite{altland1997}.
For Bose systems, a strong distinction arises between Goldstone and non-Golstone modes~\cite{gurarie2002,*gurarie2003}.
For instance, in a Bose superfluid, while localization is at its strongest at low energy for particle-like excitations,
it is suppressed for phonon excitations~\cite{bilas2006,lugan2007b,lugan2011}.
This conclusion is based on a weak disorder analysis and holds in dimension $d \leq 2$ where localization occurs for arbitrary weak disorder.
It is, however, challenged in higher dimension where the onset of the Anderson transition
requires sufficiently strong disorder, which may alter the very nature of the excitations.

In this work we study the transport of collective excitations
in a disordered, weakly-interacting Bose superfluid in dimension higher than one.
We show that the competition of disorder, screening, and density depletion
yields a strongly non-monotonic and non-universal
energy dependence of the disorder parameter,
which controls the localization properties.
In three dimensions (3D), our analysis indicates that the localization diagram exhibits
several classes of mobility spectra,
characterized by either no or several mobility edges.
We finally discuss implications of these localization properties on
quantum-quench experiments with disordered ultracold atoms.

\begin{figure}[t!]
\begin{center}
 \infigbis{25em}{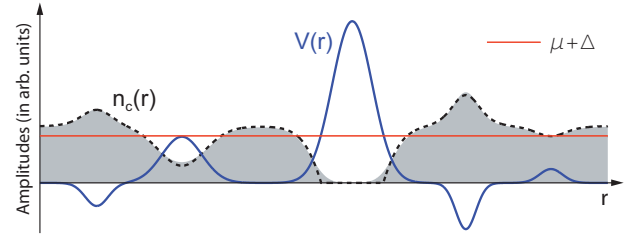}
\end{center}
\caption{\label{fig:MF}
One-dimensional cut of the density profile of a Bose superfluid in a disordered potential:
Exact numerical solution of the GPE~(\ref{eq:GPE}) [shaded area, $\nc(\textbf{r})$]
vs.\ self-consistent solution using Eq.~(\ref{eq:nc}) [dashed line],
for a disordered potential [full line, $V(\textbf{r})$]
of amplitude $\Vr=0.87\mu$ and correlation  length $\sigmar=\xi$.
The density profile follows the modulations of the potential, smoothed at the length scale of $\xi$,
and may be locally depleted around disorder maxima.
}
\end{figure}

\section{Transport theory of collective excitations}
\subsection{Meanfield scattering theory}
To study the transport of collective excitations
in the presence of disorder, it is worth devising a scattering problem.
For weakly-interacting Bose superfluids, we may rely on meanfield theory~\cite{popov1983}.
The background density field $\nc(\textbf{r})$ obeys the Gross-Pitaevskii equation (GPE),
\begin{equation}
  \big[-\hbar^{2}\nabla^2/2m  + \Vrand(\pt) - \mu + g\nc(\pt)\big]\sqrt{\nc(\pt)} = 0,
 \label{eq:GPE}
\end{equation}
where
$m$ is the particle mass,
$\mu$ is the chemical potential, and
$g>0$ is the coupling constant of short-range repulsive interactions.
The disordered potential $V(\textbf{r})$ is chosen to be spatially homogeneous, isotropic, and of vanishing statistical average. 
For weak disorder, Eq.~(\ref{eq:GPE}) can be solved perturbatively~\cite{lsp2006,gaul2008}.
Below, we consider regimes of strong disorder and it is necessary to extend this approach,
including possible local depletion of the density around the disorder maxima.
To do so, we generically write the density field in the form 
\begin{equation}
 \nc(\pt) =[\mu-\eta(\pt)+\Delta]/g,
 \label{eq:nc}
\end{equation}
where the field $\eta(\pt)$ describes the modulations of the density due to the disorder, and the quantity $\Delta$ is a shift in the chemical potential. The latter allows us to impose the conventional condition that $\eta(\pt)$ is of zero statistical average. Notice that since the density is positive everywhere, $\nc(\pt) \geq 0$, the field $\eta(\pt)$ is bounded above, $\eta(\pt)\leq \mu+\Delta$.
We then insert Eq.~(\ref{eq:nc}) into the Gross-Pitaevskii equation~(\ref{eq:GPE}), and linearize it,
which yields
\begin{equation}
 \eta(\pt)=(\mu+\Delta)~\min\left\{ \frac{\Delta+\tilde{V}(\pt)}{\mu+3\Delta/2},1 \right\},
 \label{eq:eta}
\end{equation}
where, in Fourier space,
\begin{equation}
\tilde{V}(\mathbf{q})=\dfrac{V(\mathbf{q})}{1+\xi_{\Delta}^{2}|\mathbf{q}|^{2}}.
\label{eq:ncSC1}
\end{equation}
The quantity $\tilde{V}(\pt)$ is a generalized smoothed potential~\cite{lsp2006},
where the healing length is renormalized by the shift $\Delta$
to the value $\xi_{\Delta}=\hbar/\sqrt{4m(\mu+3\Delta/2)}$.
The zero-average condition, $\langle\eta(\pt)\rangle=0$, then yields
\begin{equation}
0=\Delta+ \langle\min\{\tilde{V}(\pt),\mu+\Delta/2\}\rangle,
\label{eq:ncSC2}
\end{equation}
where $\langle \ldots \rangle$ denotes statistical averaging.
Note that Eq.~(\ref{eq:ncSC2}) ensures that $\mu+3\Delta/2\geq 0$, so that $\xi_\Delta$ is well defined.
In general, the density field $\nc(\pt)$ is therefore found by solving self-consistently
Eqs.~(\ref{eq:ncSC1}) and (\ref{eq:ncSC2}) for $\Delta$ and $\tilde{V}(\pt)$,
and Eq.~(\ref{eq:eta}) for $\eta (\pt)$.

This self-consistent solution is in good agreement with the exact numerical solution of the GPE~(\ref{eq:GPE}) (see Fig.~\ref{fig:MF}).
As expected, the density modulations $\eta(\mathbf{r})$ follow those of the disorder, smoothed at the length scale of the healing length~\cite{lsp2006,gaul2008}. However, both the amplitude of the smoothed potential and the healing length are renormalized by the energy shift $\Delta$.
Moreover, for strong disorder, the field $\eta(\pt)$ locally saturates to the constant value $\mu+\Delta$ at positions where $\tilde{V}(\pt)$ typically exceeds the chemical potential $\mu$ (more precisely, where $\tilde{V}(\pt)\geq \mu+\Delta/2$). In those regions, which will be referred to as \textit{depleted regions}, we thus have $\nc(\pt)\approx0$ (see Fig.~\ref{fig:MF}). 
In order to interpret the shift $\Delta$, we may rewrite Eq.~(\ref{eq:ncSC2}) in the form $\Delta = \int_\textrm{depl.}d\tilde{V}\,P(\tilde{V})\,[\tilde{V}-(\mu+\Delta/2)]$,
where $P(\tilde{V})$ is the probability distribution of the smoothed potential
and the integral is restricted to the depleted regions. The quantity $\Delta$ can thus be assimilated to the weight of the part of the smoothed potential that is truncated in the depleted regions.
In particular, in the case of weak disorder for which $\tilde{V}(\pt)$ never exceeds $\mu$,
we find $\Delta=0$ and we recover the solution given by usual perturbation theory~\cite{lsp2006,gaul2008}.
For stronger disorder, $\Delta$ is finite, and our approach accounts for the local depletion of the density around the disorder maxima.\\

Knowing the density field $\nc(\pt)$, we now treat the collective excitations.
The phase and density fluctuations, $\hat{\theta}$ and $\delta\hat{n}$,
are readily found by developing the many-body Hamiltonian up to order two in the operator
$\hat{B}(\pt) \equiv \delta\hat{n}(\pt)/2\sqrt{\nc(\pt)}+i\sqrt{\nc(\pt)}\hat{\theta}(\pt)$.
The resulting quadratic Hamiltonian is then diagonalized by the Bogoliubov transform
$\hat{B}(\pt) =
\sum_\varepsilon \{u_\varepsilon(\pt)\hat{b}_{\varepsilon} + v_\varepsilon^*(\pt)\hat{b}_\varepsilon^\dagger\}$,
where $\hat{b}_\varepsilon$ is the annihilation operator of an elementary pair excitation of energy $\varepsilon$. 
It yields the Bogoliubov-de Gennes equations~\cite{popov1983,degennes1995} 
\begin{eqnarray}\label{eq:BdGE}
\mathcal{L}_0 \left( \begin{matrix} u_{\varepsilon}  \\ v_{\varepsilon}  \end{matrix} \right)+ \mathcal{U}(\pt) \left( \begin{matrix} u_{\varepsilon} \\ v_{\varepsilon} \end{matrix} \right)=\varepsilon \left( \begin{matrix} u_{\varepsilon}  \\ v_{\varepsilon}  \end{matrix} \right),
\end{eqnarray}
where 
\begin{eqnarray}
\mathcal{L}_0=\left( \begin{matrix} -\dfrac{\hbar^{2}\nabla^{2}}{2m} +\mu+2\Delta & +\mu+\Delta \\ -\mu-\Delta & +\dfrac{\hbar^{2}\nabla^{2}}{2m} -\mu-2\Delta \end{matrix} \right)
\label{eq:BBdGE1}
\nonumber
\end{eqnarray}
and
\begin{eqnarray}
 \mathcal{U}(\pt) = \left( \begin{matrix} +\Vrand(\pt) -2\eta(\pt) & -\eta(\pt) \\ +\eta(\pt) & -\Vrand(\pt) +2\eta(\pt) \end{matrix} \right).
\label{eq:BBdGE2}
\nonumber
\end{eqnarray}
In this form, Eq.~(\ref{eq:BdGE}) devises a well-defined two-wave scattering problem.
The dynamics of a given excitation at energy $\varepsilon$ is governed by the homogeneous propagator $\mathcal{L}_0$ and scattering from a disordered medium defined by $\mathcal{U}(\pt)$.
The latter combines the two random fields $V(\pt)$ and $\eta(\pt)$,
which are strongly correlated [see Eqs.~(\ref{eq:eta}) to (\ref{eq:ncSC2})].

At this point, one could wonder whether the Bogoliubov approach is valid even in the presence of strong density depletion. The main approximation here is the truncation of the many-body Hamiltonian at second order in the Bogoliubov operator.
Since the latter is equivalent to the linearization of the time-dependent Gross-Pitaevskii equation (tGPE)~\cite{pitaevskii2004},
it can be tested by comparing the meanfield dynamics predicted by the exact tGPE on the one hand and by the linearized tGPE on the other hand. Our results show excellent agreement between the two in all regimes, namely the phonon, particle, intermediate regimes for weak to strong density depletion (for details, see appendix.~\ref{sec:A1}). It validates the use of the Bogoliubov approach used here.

\subsection{One-parameter scaling theory}
Universal transport properties can now be inferred using
the one-parameter scaling (OPS) approach~\cite{abrahams1979},
which can be extended to the case of excitations, as we outline here.
It consists in developing a renormalization-group (RG) analysis of the size-dependent conductance. The latter is identified to the \textit{Thouless number}~\cite{thouless1974},
which is the ratio of the energy scale associated to diffusion across a finite sample of size $L$, $\delta \varepsilon = \hbar\diffB/L^2$ (with $\diffB=w_\varepsilon\lB/d$ the classical diffusion constant, $w_\varepsilon=\hbar^{-1}\vert\partial\varepsilon/\partial\mathbf{k}\vert$ the excitation velocity, and $\lB$ the Boltzmann transport mean free path),
to the energy-level spacing, $\Delta \varepsilon = 1/N(\varepsilon)L^d$ (with $N(\varepsilon)$ the density of states per unit volume).
In diffusive regimes, if $k_\varepsilon$ is the momentum associated to the energy $\varepsilon$,
then $N(\varepsilon) \propto k_\varepsilon^{d-1}/\vert\partial\varepsilon/\partial\mathbf{k}\vert=k_\varepsilon^{d-1}/\hbar w_\varepsilon$,
so that $G(L) \propto (k_\varepsilon\lB)(k_\varepsilon L)^{d-2}$,
and $\beta \equiv d\log G / d\log L \sim d-2$.
In localized regimes, the conductance is exponentially small, $G(L) \sim \exp(-L/\Lloc)$ with $\Lloc$ the localization length, and $\beta \sim \log G$.
For $d \leq 2$, $\beta(G)$ is strictly negative and $G(L)$ always flows down to the localized regime.
Then, all states are localized, with the localization length $\Lloc \propto \lB$ in 1D
and $\log(\Lloc / \lB) \propto k_\varepsilon\lB$ in 2D.
Conversely, for $d>2$, the RG flow has an unstable fixed point at $k_{\varepsilon}\lB \sim 1$,
known as the mobility edge or the Anderson localization transition~\cite{abrahams1979}.
Since the above scaling laws are independent of the dispersion relation, these features are all universal,
except the transition point,
which is determined by the value of the inverse disorder parameter (IDP) $k_{\varepsilon}\lB$.

\subsection{Disorder parameter}
In order to estimate the IDP for the scattering problem~(\ref{eq:BdGE}), we follow the approach of Ref.~\cite{lugan2007b,lugan2011} and extend it to strong disorder with possible local density depletion (see details in appendix~\ref{sec:A2}).
In brief, we note that the homogeneous propagator $\mathcal{L}_0$ does not support only plane-wave modes with momentum $k_{\varepsilon}$ such that $\hbar^{2}k_{\varepsilon}^{2}/2m=\sqrt{\varepsilon^{2}+(\mu+\Delta)^{2}}-(\mu+2\Delta)$,
but also evanescent modes of penetration length $\gamma_{\varepsilon}^{-1}$ such that $\hbar^{2}\gamma_{\varepsilon}^{2}/2m=\sqrt{\varepsilon^{2}+(\mu+\Delta)^{2}}+(\mu+2\Delta)$. The latter ensures that for a scattering length larger than the penetration length, the excitation modes $(u_{\varepsilon},v_{\varepsilon})$ can be decomposed into two fields $(g_{\varepsilon}^{+},g_{\varepsilon}^{-})$,
where the second one
is enslaved by the first one.
Retaining only the leading disorder terms, the behavior of the excitation is then entirely determined by the field $g^{+}_{\varepsilon}$, which fulfills the closed equation
\begin{equation}
\dfrac{\hbar^{2}k_{\varepsilon}^{2}}{2m}g^{+}_{\varepsilon}(\pt)= -\dfrac{\hbar^{2}}{2m}\nabla^{2} g^{+}_{\varepsilon}(\pt) + \Vscreen(\pt) g^{+}_{\varepsilon}(\pt)
\label{eq:SchEff}
\end{equation}
where 
\begin{equation}
\Vscreen(\pt)= V(\pt)-f(\varepsilon)\eta(\pt),
\label{eq:Vscreen}
\end{equation}
with $f(\epsilon)=\frac{2\sqrt{\varepsilon^{2}+(\mu+\Delta)^{2}}-(\mu+\Delta)}{\sqrt{\varepsilon^{2}+(\mu+\Delta)^{2}}}$.
The so-called screened potential $\Vscreen(\pt)$
results from the competition of the bare disorder $V(\pt)$ and the meanfield repulsive interaction,
determined by the field $\eta(\pt)$.
This competition is strongly energy dependent due to the factor $f(\varepsilon)$.
Equation~(\ref{eq:SchEff}) describes an equivalent scattering problem, which can now be solved by standard quantum transport theory~\cite{rammer1998}.
In the on-shell approximation~\cite{kuhn2007}, it yields
\begin{equation}
\dfrac{1}{k_{\varepsilon}\lB(\varepsilon)} \simeq
\dfrac{2\pi m^{2}}{\hbar^{4}k_{\varepsilon}^{4-d}} \int\dfrac{d\Omega_{d}}{(2\pi)^{d}} (1\!-\!\cos\theta)\mathcal{C_\varepsilon}[2k_{\varepsilon}\sin (\theta/2)],
\label{eq:kelbana}
\end{equation}
where $d\Omega_{d}$ denotes the infinitesimal solid angle in $d$ dimensions
and $\mathcal{C_{\varepsilon}}(\mathbf{q}) \propto \langle \vert \Vscreen(\mathbf{q}) \vert^2 \rangle$ is the power spectrum of the screened potential. Notice that in 1D, the angular integral in Eq.~(\ref{eq:kelbana}) reduces to $\theta=\pi$ so that one recovers the result of Ref.~\cite{lugan2007b,lugan2011} for the Lyapunov exponent.

It is worth pointing out that the previous approach describes only excitations of energy $\varepsilon > \varepsilon_c \equiv \sqrt{2\Delta(\mu+3\Delta/2)}$. Otherwise, we have $k_{\varepsilon}^2<0$ and all modes of $\mathcal{L}_0$ are evanescent. In the following, we will thus disregard the case of excitations $\varepsilon < \varepsilon_c$, which in most cases reduces to a very narrow energy range at the bottom of the spectrum, since $\Delta\ll\mu$. It should as well be pointed out that since it retains only leading terms in disorder, our theory is not expected to be quantitatively exact, but rather to provide a qualitative description of the relevant physics. Possible extensions of the approach are discussed in the conclusion.

\section{Localization of Bogoliubov quasiparticles in an impurity model}
\subsection{The impurity model}

\begin{figure}[t!]
\begin{center}
 \infigbis{25em}{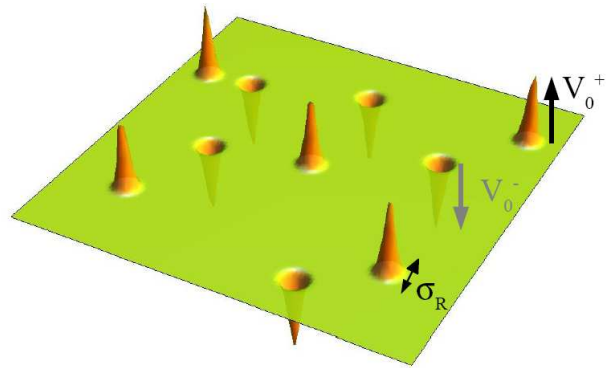}
\end{center}
\caption{\label{fig:modeleImp}
Schematic view of the two-impurity model defined by Eq.~(\ref{eq:modeleImp}).
It is made of two types of impurities with, respectively, positive ($+V_{0}^{+}$) and negative ($-V_{0}^{-}$) amplitudes.
The impurities are randomly and independently spread over in space. Each impurity is Gaussian shaped with a width $\sigmar$.
}
\end{figure}

We can now discuss the behavior of the disorder parameter of the excitations.
For concreteness, let us consider a generic impurity model described by the potential
\begin{equation}
\Vrand(\pt)=\sum_j V_{j}h(\pt-\mathbf{r}_{j})-\overline{V}.
\label{eq:modeleImp}
\end{equation}
The impurities are independent Gaussian-shaped potentials of width $\sigmar$,
$h(\pt)=\exp(-\pt^{2}/2\sigmar^{2})$,
randomly distributed in space with density $\rho$,
and with amplitudes $V_j$ that take the values $+V_{0}^{+}>0$ or $-V_{0}^{-}<0$ with equal probability (see Fig.~\ref{fig:modeleImp}).
The constant term $\overline{V}\equiv\rho(\sqrt{2\pi}\sigmar)^{d}(V_{0}^{+}-V_{0}^{-})/2$ ensures that the potential $\Vrand(\pt)$ is of vanishing statistical average.
The square disorder amplitude is $\Vr^2\equiv\langle{V^2}\rangle=\rho(\sqrt{\pi}\sigmar)^{d}[(V_{0}^{+})^2+(V_{0}^{-})^2]/2$.
This two-impurity model generalizes the one-impurity model, which is widely used in studies of Anderson localization in non-interacting systems~\cite{lifshits1988,abrahams2010}. Here, we introduce two types of impurities, namely repulsive ($V_j=+V_{0}^{+}>0$) and attractive ($V_j=-V_{0}^{-}<0$) ones. In contrast to non-interacting systems, it is crucial to distinguish repulsive and attractive impurities in the present work because they have radically different effects on the density background. For instance, only repulsive impurities can induce local density depletion.
The two-impurity model is generic in the sense that it is the simplest one to describe a disordered system where different kinds of impurities are present.
In addition, controlled impurity models can be realized in ultracold-atom systems where the impurities are made of individual atoms trapped at some random sites of an optical lattice~\cite{gavish2005,*paredes2005}. So far, only one-impurity models have been realized~\cite{gadway2011} but they can be extended to models with different kinds of impurities using different atomic species.

\subsection{Behavior of the disordered parameter}
\begin{figure}[t!]
\begin{center}
 \infigbis{22em}{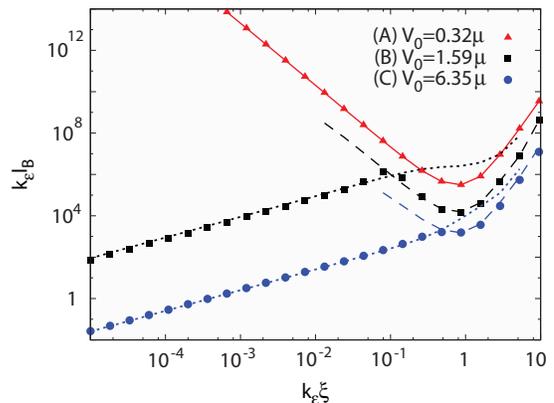}
\end{center}
\caption{
\label{fig:kelb3D}
Inverse disorder parameter (IDP) versus pair-excitation momentum for the 3D balanced impurity model, with $\sigmar=\xi$, $\rho\sigmar^3=2\times10^{-4}$, and for various values of the disorder amplitude $V_0 \equiv V_0^+=V_0^-$.
Shown are the results of Eq.~(\ref{eq:kelbana}),
where the screened power spectrum $\mathcal{C_{\varepsilon}}(\mathbf{q})$
is calculated with the full screened disorder of Eq.~(\ref{eq:Vscreen}) (dots)
or with Eq.~(\ref{eq:equation1}) (solid line),
as well as the contributions of the bulk (dashed lines)
and depleted (dotted lines) regions.
}
\end{figure}

To compute the IDP for the previous impurity model, we first numerically determine the density background $\eta (\pt)$ and $\Delta$ following the previous self-consistent procedure. This permits to compute the screened potential Eq.~(\ref{eq:Vscreen}) and its power spectrum, from which the IDP is inferred using Eq.~(\ref{eq:kelbana}).
Figure~\ref{fig:kelb3D} shows the energy dependence of the IDP
for the 3D balanced impurity case ($V_0^+=V_0^-$), plotted as a function of $k_\varepsilon\xi$.
Qualitatively similar curves are found for lower dimensions and for imbalanced impurity cases ($V_0^+ \neq V_0^-$). Depending on the disorder strength, the IDP exhibits three generic behaviors.
Notice that $k_\varepsilon\xi\rightarrow 0$ corresponds to $\varepsilon \rightarrow \varepsilon_c$.

For weak disorder (case~A in Fig.~\ref{fig:kelb3D}),
the IDP shows a non-monotonic energy dependence, which can be understood as follows.
At high energy, the excitations are insensitive to the density background
and behave as particles in the bare disorder potential.
Conversely, at low energy, the excitations are strongly affected by the
density background, which screens the disorder and suppresses scattering.
More precisely, this holds when the chemical potential exceeds the maximum of the smoothed potential,
\ie\ for $V_0^{+}\tilde{h}(0)-\overline{V}<\mu$ where $\tilde{h}(\pt)$ is the smoothed impurity.
Then, the density background as no depleted region.
The power spectrum of the screened potential, $\mathcal{C_{\varepsilon}}(\mathbf{q})$, can be
computed explicitly as a function of that of the bare disorder, $C(\mathbf{q})$,
and of the excitation energy $\varepsilon$ using Eq.~(\ref{eq:Vscreen}).
It yields
\begin{equation}
\mathcal{C_{\varepsilon}}(\mathbf{q})
=
\left(
1-\frac{1+4k_{\varepsilon}^{2}\xi^2}{1+2k_{\varepsilon}^{2}\xi^2}\frac{1}{1+\mathbf{q}^{2}\xi^{2}}
\right)^{2}
C(\mathbf{q}).
\label{eq:equation1}
\end{equation}
Inserting this Eq.~(\ref{eq:equation1}) into Eq.~(\ref{eq:kelbana}), we find the solid line in Fig.~\ref{fig:kelb3D}, which reproduces very well the data
in the full energy range for case A.
Notice that the same formula as found from Eqs.~(\ref{eq:kelbana}) and (\ref{eq:equation1}) was inferred from calculations of the scattering mean free path using a different approach in Ref.~\cite{gaul2011b}.
Equation~(\ref{eq:equation1}) generalizes the 1D case~\cite{lugan2007b,lugan2011}.
It defines a screening function [prefactor in the rhs of Eq.~(\ref{eq:equation1})],
which can also be identified in single-scattering processes~\cite{gaul2008}
and renormalizes the disorder by the interactions.
The behavior of the IDP
can now be found by inspection of Eqs.~(\ref{eq:kelbana}) and (\ref{eq:equation1}).
For $k_\varepsilon \gg \xi^{-1}, \sigmar^{-1}$, the screening is irrelevant and we can replace $\mathcal{C_{\varepsilon}}(\mathbf{q})$ by $C(\mathbf{q})$ in Eq.~(\ref{eq:kelbana}).
We then recover the free-particle behavior~\cite{kuhn2007,piraud2012a,*piraud2013b},
\begin{equation}
k_\varepsilon\lB \sim \left(\frac{\mu^2\sigmar}{\Vr^2\xi}\right) (k_\varepsilon\xi)^5,
\qquad k_\varepsilon \gg \xi^{-1}, \sigmar^{-1}.
\label{eq:klB1}
\end{equation}
Conversely, for $k_\varepsilon \ll \xi^{-1}, \sigmar^{-1}$,
the screening
strongly enhances $k_\varepsilon\lB$ compared to the free-particle case
and we find
\begin{equation}
k_\varepsilon\lB \sim \left(\frac{\mu^2\xi^d}{\Vr^2\sigmar^d}\right)(k_\varepsilon\xi)^{-d},
\qquad k_\varepsilon \ll \xi^{-1}, \sigmar^{-1}.
\label{eq:klB2}
\end{equation}
This result is in agreement with the universal behavior expected for Goldstone modes~\cite{gurarie2002,*gurarie2003}.
Both low-energy and high-energy scalings reproduce the behavior of case A and locate the minimum of the IDP
at $\kmin \sim \min (1/\xi, 1/\sigmar)$.
Note that a non-monotonic behavior of the IDP is also found in the propagation of other kinds of waves, such as
photonic~\cite{john1984} and acoustic~\cite{kirkpatrick1985} ones.

For intermediate to strong disorder (cases B and C in Fig.~\ref{fig:kelb3D}),
the energy dependence of the IDP
found from the solution of the full scattering problem [Eqs.~(\ref{eq:Vscreen}) and (\ref{eq:kelbana})]
strongly differs from the weak disorder case at low energy, where $k_\varepsilon\lB$ now increases with the energy.
To understand this, it should be noticed that, for $V_0^{+}\tilde{h}(0)-\overline{V}>\mu$,
the background density is now locally depleted around the positive impurities.
Hence, during its propagation, an excitation goes through two types of regions,
namely the \textit{density depleted} region,
and the rest, which constitutes the \textit{density bulk}.
In the bulk, the field $\eta(\pt)$ may be approximated by $\tilde{V}(\pt)$, provided we neglect the quantity $\Delta$,
which is valid for low impurity density, $\rho\sigmar^d \ll 1$.
It yields a non-monotonic contribution to $k_\varepsilon\lB$ similar to case A,
with a smaller overall magnitude due to the truncation around the positive impurities (dashed lines in Fig.~\ref{fig:kelb3D}).
Conversely, in the depleted regions, the field $\eta (\pt)$ saturates to the value $\mu+\Delta$. The bare disorder in those regions is thus protected against screening and Eq.~(\ref{eq:Vscreen}) may be replaced by $\mathcal{V}_\varepsilon(\pt) \simeq V(\pt)-(\mu+\Delta)f(\varepsilon)$.
In this field, the excitations behave as (non-Goldstone) free particles,
yielding a monotonic contribution to $k_\varepsilon\lB$
(dotted lines).
In the white-noise limit ($k_\varepsilon\sigmar \ll 1$),
this contribution is
\begin{equation}
k_\varepsilon\lB \sim (k_\varepsilon\xi)^{4-d},
\qquad k_\varepsilon \ll \sigmar^{-1}.
\label{eq:klB3}
\end{equation}

The various behaviors of the IDP observed in Fig.~\ref{fig:kelb3D}
can then be interpreted as follows.
Neglecting the correlations between the contributions of the bulk and depleted regions,
the disorder parameter $(k_\varepsilon\lB)^{-1}$ is approximately
the sum of these two contributions.
Its inverse (the IDP $k_\varepsilon\lB$, which is plotted in Fig.~\ref{fig:kelb3D}),
is thus dominated by the smallest corresponding contribution.
At low energy, because of the screening in the bulk, the contribution of the depleted region always dominates if it exists, and captures the free-particle-like behavior of $k_{\varepsilon}\lB$.
At intermediate energy the behavior of $k_\varepsilon\lB$
crucially depends on the relative magnitude of the two contributions.
When $V_0^{+}\tilde{h}(0)-\overline{V}\gtrsim\mu$ (case B), only the upper fraction of the positive impurities is truncated and the bulk starts to dominate at moderate energy. It results in a turning point $\kmax$ where bulk and depleted regions equally contribute to the IDP, yielding there a local maximum.
When $V_0^{+}$ increases, the depleted region dominates in a wider energy range and $\kmax$ moves to higher values.
When $V_0^{+}\tilde{h}(0)-\overline{V}\gg\mu$ (case C), the positive impurities are almost entirely truncated, and $\kmax$ eventually merges with the local minimum $\kmin$, which does not significantly depend on $V_0^{+}$.
The curve then becomes monotonic.

\subsection{Localization diagram}
\begin{figure}[t!]
\begin{center}
 \infigbis{24em}{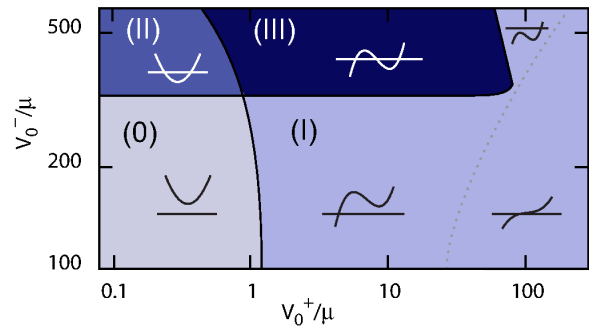}
\end{center}
\caption{Localization diagram of pair excitations in the 3D impurity model,
plotted as a function of the amplitudes of
positive ($V_0^{+}$) and negative ($V_0^{-}$) impurities,
for $\sigmar=\xi$ and $\rho\sigmar^3=2\times10^{-4}$.
It exhibits four classes of mobility spectra, characterized by
zero (0), one (I), two (II), or three (III) mobility edges.
Note the different scales on the two axes.
\label{fig:diagram}
}
\end{figure}
We now turn to the localization properties of the collective excitations,
focusing on the 3D case where mobility edges appear at the localization threshold $k_{\varepsilon}\lB \sim 1$.
We determine the latter from IDP curves as those of Fig.~\ref{fig:kelb3D}
for the general model with different amplitudes of the positive ($V_0^+$) and negative ($V_0^-$) impurities.
In all cases, they are of type A, B, or C with an overall magnitude and positions of the local maximum and minimum
that depend on the parameters of the disorder potential.
The resulting localization diagram, shown on Fig.~\ref{fig:diagram},
displays several localization classes with
between zero and three mobility edges in the excitation spectrum.
The high energy states are always extended.
The regions of the diagram are then determined by three conditions.
Firstly, the existence of depleted regions requires $V_0^+\tilde{h}(0)-\overline{V}\geq\mu$,
which defines the roughly vertical line on the diagram.
On the left, the IDP curves are of type A and the low-energy excitations are extended ($k_\varepsilon\lB > 1$).
Just on the right, they are of type B. The low-energy excitations are then localized ($k_\varepsilon\lB < 1$),
and there is at least one mobility edge.
Secondly, when the local minimum of the IDP curve,  $(k_\varepsilon\lB)^\textrm{min}$,
is below the localization threshold a band of localized states appear at intermediate energy,
giving rise to two additional mobility edges.
For $V_0^+ \ll V_0^-$, the condition reads $1 \sim (k_\varepsilon\lB)^\textrm{min} \propto 1/(V_0^-)^2$,
which yields the nearly horizontal line on the diagram.
Thirdly, when $V_0^+$ increases, the local maximum of the IDP
curve decreases. In the region with three mobility edges (III), the two low-energy ones disappear.
In the region with one mobility edge (I), the IDP curve turn from type B to type C,
without affecting the number of mobility edges.

The localization diagram of Fig.~\ref{fig:diagram} is expected to be generic.
In particular, the competition of disorder, screening, and density depletion
determine the diversity of mobility spectra.
Yet, a given model of disorder does not necessary display all cases
and the imbalanced impurity model with a finite correlation length
is the simplest we found that does.
For instance, for only positive impurities or in the balanced case,
the only possibilities are (0) or (I)
because the minimum of the IDP cannot be controlled independently of the density depletion.
Conversely, for only negative impurities, the depleted region is absent and
the only possibilities are (0) or (II).
The case of white-noise disorder is also limited because the smoothed impurity potential diverges in the center,
$\tilde{h}(\pt)=\e^{-r/\xi}/4\pi\xi^{2}r$,
so that depleted regions strictly appear as soon as $V_0^+ \neq 0$,
and the only possibilities left are (I) and (III).

\section{Conclusions}
The physics we have discussed here is particularly relevant to ultracold-atom experiments.
In these systems, out-of-equilibrium dynamics can be generated by a local quench, which produces collective excitations~\cite{lauchli2008,*cheneau2012,*trotzky2012,*langen2013,*carleo2014}. In the presence of disorder, their transport properties and ability to mediate long-range energy transfer are determined by the four classes of mobility spectra of the localization diagram. 
In case~(0), all excitations are protected against localization and propagate diffusively,
\ie\ $\langle r^2 \rangle = 2 D_\textrm{\tiny B} t $.
In all other cases, energy can only be partially transferred since some excitation modes are localized. 
Energy-resolved quenches may provide experimental evidence of such mobility spectra in ultracold gases. 
Moreover, these systems offer a wide range of models of disorder,
\eg\ impurities~\cite{gavish2005,*paredes2005,gadway2011}
and speckle potentials~\cite{lsp2010,*modugno2010,*shapiro2012}.
The statistical properties of the latter may be tailored,
which may lead to even richer localization diagrams~\cite{plodzien2011,piraud2012b,piraud2012a}.

The observation of the localization effects we have discussed here requires that the lifetime $\tau$ of the Bogoliubov quasiparticles exceeds the transport meanfree time $\tau_\textrm{\tiny B} \equiv l_\textrm{\tiny B}/w_\varepsilon$ (with $w_\varepsilon$ the excitation group velocity).
On the one hand, for low temperature, the decay of Bogoliubov excitations is dominated by Beliaev processes~\cite{beliaev1958a,*beliaev1958b,pitaevskii2004}.
To estimate the corresponding decay rate $\Gamma=1/\tau$, we resort to local density approximation.
The depleted regions, where the excitations behave as free particles with infinite lifetime,
very weakly contribute to $\Gamma$.
A good estimate of $\Gamma$ is thus given by the bulk contribution.
For a typical excitation $\varepsilon\sim\mu$, it yields
$\tau \simeq m/10\hbar n^{3/2} a_{\textrm{sc}}^{5/2}$,
where $a_{\textrm{\tiny sc}}=mg/4\pi\hbar^2$ is the scattering length~\cite{beliaev1958a,*beliaev1958b}.
On the other hand, $w_\varepsilon \simeq \sqrt{gn/m}$ in the phonon regime and $l_\textrm{\tiny B} \sim 1/k_\varepsilon$ in the region of interest, which yields
$\tau_\textrm{\tiny B} \sim m/4\pi\hbar na_{\textrm{sc}}$.
Therefore, the validity condition of our approach reads $\tau_{\textrm{\tiny B}}/\tau \sim \sqrt{na_{\textrm{\tiny sc}}} \ll 1$, which is the validity condition of the Bogoliubov approach, very verified in dilute-gas Bose-Einstein condensates~\cite{pitaevskii2004}. For instance, using the parameters of Ref.~\cite{jendrzejewski2012}, we find find
$\tau \sim 6$s and $\tau_{\textrm{\tiny B}} \sim 5$ms,
making experimental observation of our predictions possible. 

The approach used in this work provides a intuitive understanding of the physics at stake as well as generic qualitative predictions for the localization behavior of collective excitations. However, since it relies on lowest-order perturbation theory, it is not expected to be quantitatively accurate.
In particular, it does not take into account the disorder-induced shift of the dispersion relation, which is known, in the non-interacting case, to result in a possibly significant shift on the position of the mobility edge~\cite{piraud2012a,piraud2013b,piraud2014}.
Moreover, numerical calculations in the non-interacting case show that the mobility edge significantly depends on the model of disorder~\cite{delande2014,fratini2015}.
Therefore, the determination of the precise localization diagram for collective pair excitations in disordered Bose superfluids, as well as the identification of the various classes of mobility spectra predicted in the present work, require a full numerical resolution of the localization problem in the two-impurity model as well as in other models of disorder.

This research was supported by
the European Research Council (FP7/2007-2013 Grant Agreement No.\ 256294),
the Minist\`ere de l'Enseignement Sup\'erieur et de la Recherche,
and
the Institut Francilien de Recherche sur les Atomes Froids (IFRAF).
We acknowledge the use of the computing facility cluster GMPCS of the
LUMAT federation (FR LUMAT 2764).


%


\bigskip

\begin{appendix}

\section{Validity of the second-order development in the presence of strong density depletion}
\label{sec:A1}

In this appendix, we show that the Bogoliubov approach used in this paper, \ie~a development of the many-body Hamiltonian around the inhomogeneous density background up to second order in fluctuations terms, is valid even in the presence of strong local density depletion.
We perform this check by comparing exact calculations and linearized theory at the meanfield level.
We consider the time-dependent Gross-Pitaevskii equation (tGPE),
\begin{equation}
 i\hbar\partial_t \psi = -\dfrac{\hbar^{2}\nabla^2}{2m}\psi + \Vrand(\pt)  + g|\psi|^2 \psi,
 \label{eq:tGPE}
\end{equation}
which governs the time-evolution of a condensate wavefunction $\psi(\pt,t)$.
In the linearized approach, one writes $\psi(\pt,t)=\sqrt{\nc(\pt)} +\delta\psi(\pt,t)$, 
where $\nc(\pt)$ is the density background found from the solution of the stationary GPE~(\ref{eq:GPE})
and $\delta\psi(\pt,t)$ is a small perturbation. At lowest order, it yields the linearized equation
\begin{eqnarray}\label{eq:lintGPE}
i\hbar\partial_t \left( \begin{matrix} \delta\psi  \\ \delta\psi^* \end{matrix} \right)=\mathcal{L}_{GP}\left( \begin{matrix} \delta\psi  \\ \delta\psi^*\end{matrix} \right),
\end{eqnarray}
where the matrix
\begin{eqnarray}
\mathcal{L}_{GP}=\left( \begin{matrix}  -\dfrac{\hbar^{2}\nabla^2}{2m} + \Vrand  + g\nc-\mu & g\nc \\ -g\nc & +\dfrac{\hbar^{2}\nabla^2}{2m} - \Vrand - g\nc+\mu \end{matrix} \right)\nonumber
\end{eqnarray}
is exactly the one appearing in the Bogoliubov-de-Gennes equations~\cite{pitaevskii2004}. This linearization procedure thus turns out to be equivalent to the Bogoliubov development of the many-body Hamiltonian to second order.
Therefore, to check the validity of the latter in the presence of density depletion, one can compare the results of the time-evolution of an excitation $\delta\psi(\pt,t)$ on top of a depleted condensate $\nc(\pt)$, using either the exact tGPE~(\ref{eq:tGPE}) or its linearized version, Eq.~(\ref{eq:lintGPE}).

\begin{figure}[t!]
\begin{center}
 \infigbis{24em}{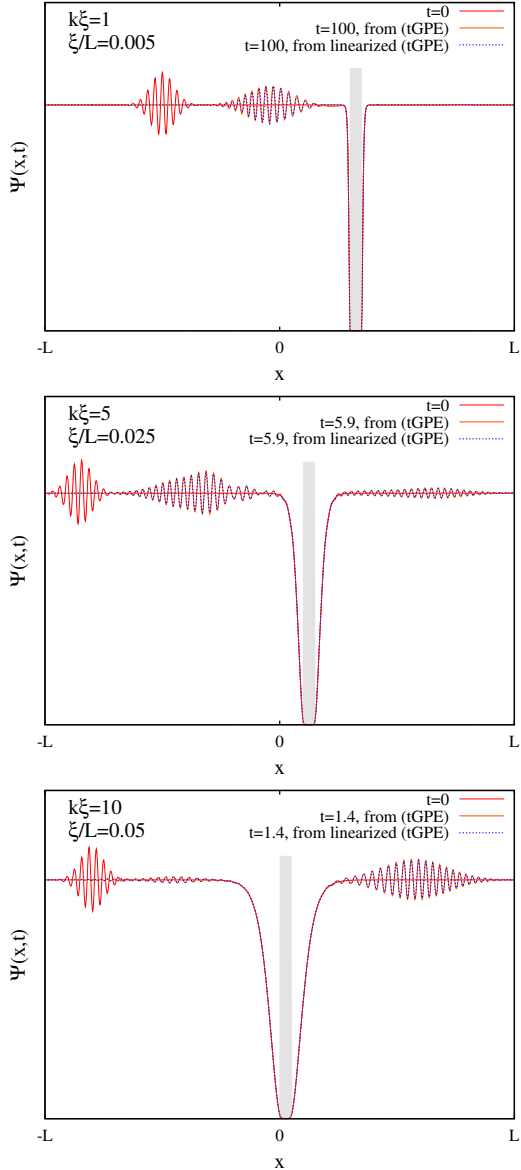}
\end{center}
\caption{Time evolution of an excitation $\delta\psi_0(x)=e^{ikx}e^{-(x-x_0)^2/2\sigma^2}$ on top of a depleted condensate in the presence of a strong potential barrier $\Vrand(x)=V_0 \Theta(a-|x-x_*|)$, for $V_0=30\mu$, $2a=0.05L$, and $\sigma=0.05L$. Three different regimes of the excitation spectrum  are considered:
(i)~$k\xi=1$ with almost total reflection (top),
(ii)~$k\xi=5$ with reflection and transmission of the excitation of the same order to magnitude (central),
(iii)~$k\xi=10$ with almost total transmission (bottom). Note that for clarity purposes, $\xi$ is varied from a panel to another.
Shown are the initial wavefunction $\psi(x,0)=\sqrt{\nc(x)}+\delta\psi_0(x)$ (solid red line)
as well as the final wavefunction as given by the full time-dependent tGPE (solid orange line)
and by the linearized tGPE (blue dotted line). The agreement between the two calculations is excellent in all cases.
\label{fig:propag}
}
\end{figure}

We have performed this test in one dimension, which is the most unfavorable dimension due to large fluctuations.
For the sake of simplicity, the density depletion is induced by a single strong potential barrier $\Vrand(x)=V_0 \Theta(a-|x-x_*|)$, of height $V_0=30\mu$, width $2a=0.05L$, and centered on some position $x_*$. The quantity $\Theta$ denotes the Heaviside function. In such a configuration, we first determine the density background $\nc(x)$ solving the stationary GPE Eq.~(\ref{eq:GPE}) by imaginary time propagation. The latter is strongly depleted under the barrier (see Fig.~\ref{fig:propag}). We then add a small excitation on top on this background, $\delta\psi_0(x)=e^{ikx}e^{-(x-x_0)^2/2\sigma^2}$. The latter describes a plane wave of momentum $k$ inside a Gaussian envelop of width $\sigma$. In practice, we choose $k\sigma\gg 1$ so that it is sufficiently narrow around $k$ in momentum space. We then compute the time evolution of this initial excitation either solving Eq.~(\ref{eq:tGPE}) for  $\psi(x,t)$ with the initial condition  $\psi(x,0)=\sqrt{\nc(x)}+\delta\psi_0(x)$, or solving Eq.~(\ref{eq:lintGPE}) for  $\delta\psi(x,t)$ with the initial condition  $\delta\psi(x,0)=\delta\psi_0(x)$. We have performed this comparison in a wide range of parameters, from non-depleted to strongly depleted cases, and from the phonon ($k\xi\ll 1$) to the free-particle ($k\xi\gg 1$) regimes. As shown on Fig.~\ref{fig:propag} in the case of strong depletion, we found an excellent agreement in all cases, irrespective to the values of $k$ and to the relative strength of reflection and transmission by the barrier. This validates the use of the linearized equation, and thus of the Bogoliubov approach, to study the dynamics of the collective excitations even in the presence of strong modulations of the potential.

\section{Derivation of the inverse disorder parameter}
\label{sec:A2}
The background density field $\nc(\pt)$ being given by the solution of Eqs.~(\ref{eq:nc}) to (\ref{eq:ncSC2}),
the collective excitations $(u_\varepsilon, v_\varepsilon)$ can now be determined by solving the Bogoliubov-de Gennes equations~(\ref{eq:BdGE}),
\begin{eqnarray}\label{eq:BdGEsup}
\mathcal{L}_0 \left( \begin{matrix} u_{\varepsilon}  \\ v_{\varepsilon}  \end{matrix} \right)+ \mathcal{U}(\pt) \left( \begin{matrix} u_{\varepsilon} \\ v_{\varepsilon} \end{matrix} \right)=\varepsilon \left( \begin{matrix} u_{\varepsilon}  \\ v_{\varepsilon}  \end{matrix} \right),
\end{eqnarray}
where
\begin{eqnarray}
\mathcal{L}_0=\left( \begin{matrix} {-\hbar^{2}\nabla^{2}}/{2m} +\mu+2\Delta & +\mu+\Delta \\ -\mu-\Delta & {+\hbar^{2}\nabla^{2}}/{2m} -\mu-2\Delta \end{matrix} \right)
\label{eq:BdGE1sup}
\nonumber
\end{eqnarray}
and
\begin{eqnarray}
 \mathcal{U}(\pt) = \left( \begin{matrix} +\Vrand(\pt) -2\eta(\pt) & -\eta(\pt) \\ +\eta(\pt) & -\Vrand(\pt) +2\eta(\pt) \end{matrix} \right).
\label{eq:BdGE2sup}
\nonumber
\end{eqnarray}
The dynamics of a given excitation at energy $\varepsilon$ is thus governed by the homogeneous propagator $\mathcal{L}_0$ and scattering from the disordered medium defined by $\mathcal{U}(\pt)$. 

In order to solve the Bogoliubov-de Gennes equations (BdGEs),
we generalize the approach of Ref.~\cite{lugan2007b} to the strong disorder case where $\Delta \neq 0$.
We first rewrite the BdGEs~(\ref{eq:BdGEsup}) in the form
\begin{eqnarray}
\dfrac{\hbar^2}{2m}\nabla^2 \left( \begin{matrix} u_\varepsilon \\ v_\varepsilon  \end{matrix} \right) & = &
\left( \begin{matrix} -\varepsilon+\mu+2\Delta & \mu+\Delta \\ \mu + \Delta & \varepsilon+\mu+2\Delta \end{matrix} \right) \left( \begin{matrix} u_\varepsilon \\ v_\varepsilon  \end{matrix} \right) \label{eq:BdGEf}\\
& & + \left( \begin{matrix} \Vrand(\pt)-2\eta(\pt) & -\eta(\pt) \\ -\eta(\pt) & \Vrand(\pt)-2\eta(\pt) \end{matrix} \right) \left( \begin{matrix} u_\varepsilon \\ v_\varepsilon  \end{matrix} \right). \nonumber
\end{eqnarray}
A suitable basis to perform diagrammatic expansion in leading disorder terms is found by applying the linear transform $(u_\varepsilon, v_\varepsilon)\rightarrow (g_\varepsilon^{+},g_\varepsilon^{-})$
that diagonalizes the homogeneous term in Eq.~(\ref{eq:BdGEf}),
\ie\ the matrix
$M \equiv \left( \begin{matrix} -\varepsilon+\mu+2\Delta & \mu+\Delta \\ \mu + \Delta & \varepsilon+\mu+2\Delta \end{matrix} \right)$.
It yields
\begin{eqnarray}
\left( \begin{matrix}  g^{+}_{\varepsilon} \\ g^{-}_{\varepsilon}  \end{matrix} \right)
\! = \!  \left( \begin{matrix}
\dfrac{\hbar^{2}\gamma_{\varepsilon}^{2}}{2m}\! - \! \Delta \! + \! \varepsilon
&
-\dfrac{\hbar^{2}\gamma_{\varepsilon}^{2}}{2m} \! + \! \Delta \! + \! \varepsilon
\\
\dfrac{\hbar^{2}k_{\varepsilon}^{2}}{2m} \! + \! \Delta \! - \! \varepsilon
&
-\dfrac{\hbar^{2}k_{\varepsilon}^{2}}{2m} \! - \! \Delta \! - \! \varepsilon
\end{matrix} \right)
\!\!
\left( \begin{matrix}  u_{\varepsilon} \\ v_{\varepsilon}  \end{matrix} \right),
\end{eqnarray}
where
$-\hbar^{2}k_{\varepsilon}^{2}/2m \equiv -\sqrt{\varepsilon^{2}+(\mu+\Delta)^{2}}+(\mu+2\Delta)$
and
$\hbar^{2}\gamma_{\varepsilon}^{2}/2m \equiv \sqrt{\varepsilon^{2}+(\mu+\Delta)^{2}}+(\mu+2\Delta)$
are the eigenvalues of the homogeneous matrix $M$.
Without any approximation at this stage, the BdGEs in the $(g_\varepsilon^{+},g_\varepsilon^{-})$ basis then read
\begin{eqnarray}
\dfrac{\hbar^{2}k_{\varepsilon}^{2}}{2m}g^{+}_{\varepsilon}(\pt) & = & -\dfrac{\hbar^{2}}{2m}\nabla^{2} g^{+}_{\varepsilon}(\pt) + \Big[ V(\pt)- f_{-}(\epsilon) \eta(\pt) \Big] g^{+}_{\varepsilon}(\pt) \nonumber\\
& & + \Phi_{+}(\epsilon) \eta(\pt) g^{-}_{\varepsilon}(\pt)
\label{eq:BdGEg+}\\
-\dfrac{\hbar^{2}\beta_{\varepsilon}^{2}}{2m}g^{-}_{\varepsilon}(\pt) & = & -\dfrac{\hbar^{2}}{2m}\nabla^{2} g^{+}_{\varepsilon}(\pt) + \Big[ V(\pt)- f_{+}(\epsilon) \eta(\pt) \Big] g^{-}_{\varepsilon}(\pt) \nonumber\\
& & + \Phi_{-}(\epsilon) \eta(\pt) g^{+}_{\varepsilon}(\pt),
\label{eq:BdGEg-}
\end{eqnarray}
with $f_{\pm}(\epsilon)=\frac{2\sqrt{\varepsilon^{2}+(\mu+\Delta)^{2}}\pm(\mu+\Delta)}{\sqrt{\varepsilon^{2}+(\mu+\Delta)^{2}}}$ and
$\Phi_{\pm}(\epsilon)=\frac{\sqrt{\varepsilon^{2}+(\mu+\Delta)^{2}}\pm(\mu+\Delta)}{\sqrt{\varepsilon^{2}+(\mu+\Delta)^{2}}}$.
In the absence of disorder,
Eqs.~(\ref{eq:BdGEg+}) and (\ref{eq:BdGEg-}) are now decoupled and are straightforward to solve.
The $g^{+}_{\varepsilon}$ modes are plane waves of momentum $k_{\varepsilon}$,
while the $g^{-}_{\varepsilon}$ are evanescent waves of penetration length $\gamma_{\varepsilon}^{-1}$.
The latter vanish identically if the system is infinite or has periodic boundary conditions.

In the presence of disorder,
we can therefore make the assumption $|g^{-}_{\varepsilon}| \ll |g^{+}_{\varepsilon}|$
since $g^{-}_{\varepsilon}$ is at least one order of magnitude smaller that $g^{+}_{\varepsilon}$ in $\Vr/\mu$ and $\Delta/\mu$. Keeping only the leading-order terms in Eq.~(\ref{eq:BdGEg-}), we then find
\begin{equation}
g^{-}_{\varepsilon} \simeq -\dfrac{2m}{\hbar^2 \gamma_\varepsilon^2} \Phi_{-}(\varepsilon) \int d\pt^\prime\ G_{1/\gamma_\varepsilon}(\pt-\pt^\prime)\eta(\pt^\prime)g^{+}_{\varepsilon}(\pt^\prime),
\label{eq:slave}
\end{equation}
where $G_{1/\gamma_\varepsilon}$ is the Green function associated to the differential operator $-\nabla^2/\gamma^2_\varepsilon +1$. In Fourier space, it reads $G_{1/\beta_\varepsilon}(\mathbf{q})=(2\pi)^{-d/2}/[1+q^2/\beta_\varepsilon^2]$ and in real space it is a positive function of integral unity, decaying exponentially on a length scale $1/\gamma_\varepsilon$.
Then, since ${2m}/{\hbar^2 \gamma_\varepsilon^2}<\mu$ and $\vert \Phi_{-}(\varepsilon) \vert <1$ for any energy $\varepsilon$, we find $$|g^{-}_{\varepsilon}|\lesssim  \int G_{1/\gamma_\varepsilon}(\pt-\pt')|\eta(\pt')||g^{+}_{\varepsilon}(\pt')|d\pt'/\mu \lesssim (\Vr/\mu) |g^{+}_{\varepsilon}|,$$ which is consistent with the working assumption $|g^{-}_{\varepsilon}| \ll |g^{+}_{\varepsilon}|$.
Therefore, the last term of Eq.~(\ref{eq:BdGEg+}) can be neglected, which yields a closed equation for $g^{+}_{\varepsilon}$,
\begin{equation}
\dfrac{\hbar^{2}k_{\varepsilon}^{2}}{2m}g^{+}_{\varepsilon}(\pt)= -\dfrac{\hbar^{2}}{2m}\nabla^{2} g^{+}_{\varepsilon}(\pt) + \Vscreen(\pt) g^{+}_{\varepsilon}(\pt),
\label{eq:Schrodeff}
\end{equation}
where
\begin{equation}
\Vscreen(\pt)= V(\pt)-f(\varepsilon)\eta(\pt)
\label{eq:Vrond}
\end{equation}
and, for simplicity, we now denote by $f(\epsilon)$ the quantity  $f_{-}(\epsilon)=\frac{2\sqrt{\varepsilon^{2}+(\mu+\Delta)^{2}}-(\mu+\Delta)}{\sqrt{\varepsilon^{2}+(\mu+\Delta)^{2}}}$.
The quantity $\Vscreen(\pt)$ defines a so-called screened potential of zero-average. It can be viewed as the screening of the bare potential $V(\pt)$ by the density background encoded in $\eta(\pt)$. 
It notably depends on the energy $\varepsilon$ of the Bogoliubov excitation.

Therefore Eq.~(\ref{eq:Schrodeff}), together with Eq.~(\ref{eq:Vrond}), contains the leading disorder terms. It features an effective single-wave scattering problem, which can now be solved by standard quantum transport theory~\cite{rammer1998}.
Localization properties are then determined in a two-step process~\cite{vollhardt1980a,*vollhardt1980b}.
Firstly, the transport meanfree path is calculated in the semi-classical approach where interference of multiple-scattering paths is neglected. Within the \textit{on-shell} approximation, which amounts to assimilate the spectral function to the disorder-free one, diagrammatic theory yields Eq.~(\ref{eq:kelbana}),
\begin{equation}
\dfrac{1}{k_{\varepsilon}\lB(\varepsilon)} \simeq
\dfrac{2\pi m^{2}}{\hbar^{4}k_{\varepsilon}^{4-d}} \int\dfrac{d\Omega_{d}}{(2\pi)^{d}} (1\!-\!\cos\theta)\mathcal{C_\varepsilon}[2k_{\varepsilon}\sin (\theta/2)],
\label{eq:kelbsuppl}
\end{equation}
for models of disorder with isotropic correlation functions~\cite{kuhn2007},
as considered in this work.
For extension to anisotropic correlation functions, see Ref.~\cite{piraud2012a,*piraud2013b}.
Secondly, localization is found using either the one-parameter scaling theory~\cite{abrahams1979}
or the self-consistent approach~\cite{vollhardt1980a,*vollhardt1980b}.
The one-parameter scaling theory is used and discussed in the main text.
The self-consistent theory incorporates interference corrections on the top of diffusive dynamics,
which yields a self-consistent equation for the diffusive constant or the localization length.
Both approaches give the approximate localization threshold $k_\varepsilon l_{\textrm{\tiny B}} \sim 1$ used in the present paper.
\end{appendix}

\end{document}